\documentstyle[twoside,fleqn,espcrc2]{article}
\include{epsf}

\newcommand{\AmS}{{\protect\the\textfont2
  A\kern-.1667em\lower.5ex\hbox{M}\kern-.125emS}}

\title{Analyzing the spectrum of general, non-hermitian Dirac operators}

\author{Christof Gattringer\address{Massachusetts Institute of Technology, 
        Center for Theoretical Physics, \\ 
        77 Massachusetts Avenue, Cambridge MA 02139 USA}
                and 
        Ivan Hip\address{Institut f\"ur theoretische Physik, 
        Karl-Franzens Universit\"at Graz \\
        Universit\"atsplatz 5, 8010 Graz, Austria}
        \thanks{Supported by Fonds zur F\"orderung der 
        wissen\-schaft\-li\-chen Forschung, Projects P11502-PHY, J1577-PHY}}       
\begin{document}

\begin{abstract}
We discuss the computational problems when analyzing general,
non-hermitian matrices and in particular the un-modified Wilson
lattice Dirac 
operator. We report on our experiences with the Implicitly Restarted
Arnoldi Method. The eigenstates of the Wilson-Dirac operator 
which have real eigenvalues and correspond to zero modes in the continuum 
are analyzed by correlating the size of the eigenvalues with the chirality
of the eigenstates. 
\end{abstract}

\maketitle

\section{THE PROBLEM}

Analyzing the eigensystem of large matrices such as the 
Dirac operator in 4-D lattice field theory is a quite demanding 
numerical task. Furthermore the lattice Dirac operator in 
Wilson formulation is neither hermitian nor anti-hermitian 
which poses an additional difficulty compared to a case with
symmetry. 

The Wilson-Dirac operator consists of a trivial part (proportional 
to the unit matrix) which comes from mass- and Wilson terms and a 
non-trivial part $K$. Here we analyze the clover-improved \cite{ShWo85}
Wilson-Dirac operator and $K$ consists of two parts $ K = Q - c_{sw} C$ 
where $Q$ is the standard Wilson hopping matrix and $C$ the clover term.
The matrix $K$ is neither hermitian nor anti-hermitian, but hermitian 
conjugation is implemented through the similarity transformation
\begin{equation}
\gamma_5 K \gamma_5 \; = \; K^\dagger \; .
\label{hercon}
\end{equation}
The relation (\ref{hercon}) allows to define the hermitian modification
$\gamma_5 K$ and several numerical studies of this modified
matrix can be found in the literature. For a deeper understanding 
of the lattice Dirac operator however one would also like to study the
original problem.

\section{THE IMPLICITLY RESTARTED ARNOLDI METHOD}

The two main techniques to solve large scale eigenvalue problems
are Krylov subspace methods:
The Lanczos algorithm, which transforms the original
matrix to tridiagonal form, or, in the case of non-hermitian matrices,
the Arnoldi method, which transforms to Hessenberg form.
Since typically the Hessenberg form is dense
it is necessary to compress
the relevant information in a Krylov subspace of small dimension.
This can be done by using the Implicitly
Restarted Arnoldi Method (IRAM) \cite{So92}.
IRAM has several favorable features: The numerical precision 
is comparable to dense methods, IRAM 
is able to find degeneracies and
it gives both eigenvalues and eigenvectors. The 
software package 
ARPACK with an efficient and reliable implementation of 
the IRAM was
developed by Sorensen and collaborators \cite{So98}.
The method turns out to be 
particularly useful when analyzing the edges of the spectrum.
It can e.g.~be set up such that the eigenvalues with 
largest real parts are computed first. This gives the eigenvalues 
in the physical branch of the spectrum. 
\begin{figure}[htp]
\epsfysize=5.6cm
\epsfbox[62 71 507 433] {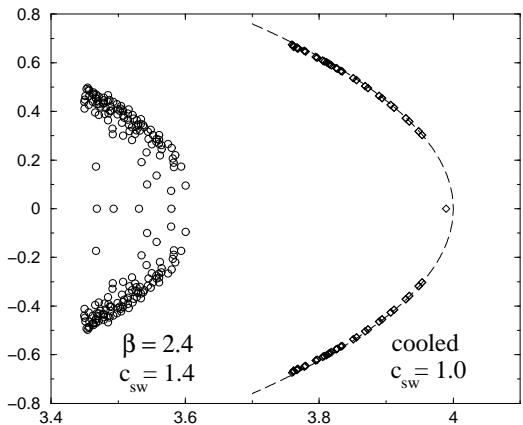} 
\vspace*{-8mm}
\caption{Physical branch of the spectrum of $K$ for a thermalized 
configuration and its cooled counterpart.}
\label{spectrum}
\vspace*{-5mm}
\end{figure}
In Fig.~\ref{spectrum}
we show the physical branch of the spectrum in the complex plane
for a thermalized SU(2) gauge field configuration
($12^4, \beta = 2.4, c_{sw} = 1.4$) and compare
it to its cooled counterpart ($c_{sw} = 1$). 
Both gauge field configurations were taken from \cite{FoGaSt97}.

As is obvious from the plot cooling leads to a strong shift
of the eigenvalues and orders them 
along a single curve (ellipse). What is also interesting
to note is the fact that the thermalized configuration 
has considerably more real eigenvalues in the physical branch than 
its cooled counterpart. In \cite{GaHi98} it was shown, that these additional 
real eigenvalues are partly due to the clover term.
This proliferation of real eigenvalues makes a probabilistic interpretation
of the Atiyah-Singer index theorem for thermalized configurations problematic,
as will be discussed in the next section.

For a complete understanding of the properties of the lattice Dirac operator
it is also interesting to analyze eigenvalues in the interior of the spectrum,
in particular on the real axis. However,
as with all Krylov subspace methods, the convergence of the interior
eigenvalues is very slow and one has to use some kind of spectral 
preconditioning. Traditionally, to find the eigenvalues in the vicinity
of some point $z$ in the complex plane a shift of the origin 
is combined 
with inversion, i.e.~the eigenvalues of
$(K - z$1\hspace{-1mm}I$)^{-1}$
are searched for. This requires solving a large linear system at
every iteration which makes this approach quite inefficient.
However, for matrices which satisfy (\ref{hercon}) a kind of sophisticated
preconditioning can be used \cite{ItIwYo87}. The
shift of the origin is combined with a multiplication with $\gamma_5$
to transform the original non-hermitian to a hermitian eigenvalue problem.
Although the explicit relation between eigenvalues of original and
transformed matrix is not known the zero eigenvalues and their eigenvectors
are the same. By shifting the origin along the real axis, it is possible
to identify real eigenvalues. In particular one traces the flow of low 
lying eigenvalues of 
$H(\rho) \equiv \gamma_5 [ K - \rho \mbox{1\hspace{-1mm}I}]$
as a function of $\rho$. Whenever the flow crosses zero at some $\rho = r$,
then this $r$ is a real eigenvalue of $K$. The chirality of the mode can be
computed from the slope of the eigenvalue flow at the crossing.
Even for this hermitian problem we found the IRAM method superior
(numerical stability,
precision and efficiency) to our implementation of the accelerated conjugate
gradient algorithm \cite{KaSi96} 
which is usually used for this purpose.
Further gain in efficiency can be achieved by using Chebyshev
polynomials to project out the desired part of the spectra.

\section{SOME RESULTS ON THE REAL PART OF THE SPECTRUM}
It was already remarked that the real eigenvalues of the lattice Dirac operator
are of particular interest. This is a consequence of (\ref{hercon})
which implies that $\psi^\dagger \gamma_5 \psi$ is different from 0 only
for eigenvectors $\psi$ of the Dirac operator which have real eigenvalues.
Thus a comparison of the chiral properties of the eigenstates implies that
only eigenvectors with real eigenvalues can play the role of the zero modes in 
the continuum \cite{ItIwYo87,NaNe94}. Based on this interpretation of the 
eigenvectors with real 
eigenvalues as the lattice zero modes one can formulate a 
lattice version of the Atiyah-Singer index theorem:
$\nu[U] = R_-  -  R_+ $.
Here $R_+$ and $R_-$ are the numbers of real 
eigenvalues in the physical branch of the spectrum with 
positive and negative chirality. The chirality is defined 
as the sign of the pseudoscalar matrix element of the corresponding 
eigenvectors and $\nu[U]$ denotes the 
topological charge of the gauge field.

The lattice version of the index theorem is expected to
hold for sufficiently smooth gauge field configurations. This was 
demonstrated for smooth SU(2) and SU(3) background configurations \cite{smooth}
and for fully quantized QED$_2$ \cite{qed2}. For thermalized SU(2) and SU(3)
configurations the situation is less clear \cite{GaHi98,ItIwYo87,NaVr97}.
There is no one-to-one corespondence \cite{GaHi98} between the value of
the topological charge assigned using the index theorem and the value from
improved cooling \cite{FoGaSt97} and only averaging over larger samples or studying 
density functions of eigenvalues \cite{NaVr97} allows a probabilistic
interpretation of the index theorem.

\begin{figure}[htb]
\epsfysize=5.3cm
\epsfbox[31 64 523 438] {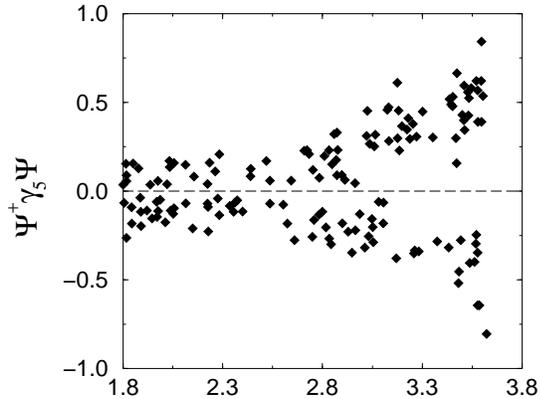} 
\vspace*{-8mm}
\caption{Correlation of the real eigenvalues with 
the corresponding pseudoscalar matrix elements.}
\label{correlation}
\vspace*{-5mm}
\end{figure}

Here\footnote{We 
thank P. de Forcrand 
for suggesting to analyze the correlation 
with the pseudoscalar matrix elements.} 
we correlate the position of the real eigenvalues with the 
size of the pseudoscalar matrix element $\psi^\dagger \gamma_5 \psi$ for
the corresponding eigenvectors $\psi$. For smooth gauge field
configurations the real eigenvalues in the physical
branch of $K$ are in the vicinity of 4 on the real axis and
the pseudoscalar matrix elements are in the vicinity of $\pm 1$, the sign given 
by the chirality of the mode. For thermalized configurations the absolute 
value of the pseudoscalar matrix elements becomes smaller and additional pairs
of real eigenvalues with opposite chirality are created \cite{GaHi98}. We check 
if the artificial real eigenvalues from the clover term have smaller 
pseudoscalar matrix elements and if this can be used to discriminate between
the would-be zero modes on the lattice and the additional real modes.
In Fig.~\ref{correlation} we show a scatter plot containing the eigenvalues 
between 1.8 and 4 on the real axis using a sample of 10 thermalized gauge field 
configurations ($12^4, \beta = 2.4, c_{sw} = 1.4$). The horizontal axis gives 
their position on the real axis and on the vertical axis we plot the pseudoscalar
matrix elements of the corresponding eigenvectors.

If there was a clear separation of the 
lattice would-be zero modes from artificial real modes
(due to strongly fluctuating fields and the clover term)
one would find two well isolated islands of points 
in the upper and lower right corners of our
plot. However, as is clear from the figure no such 
separated islands exist and we have to close with the
conclusion of \cite{GaHi98} that for $12^4, \beta = 2.4, 
c_{sw} = 1.4$ there is no proper separation between 
physical zero modes and artifacts.

\end{document}